\pdfoutput=1
\documentclass[%
reprint,
superscriptaddress,
nofootinbib,
 amsmath,amssymb,
 aps,
 prx,
 longbibliography,
floatfix,
]{revtex4-1}

\usepackage{graphicx}
\usepackage{dcolumn}
\usepackage{bm}
\usepackage{appendix}
\usepackage{hyperref}
\usepackage{graphicx}
\usepackage{amsmath}
\usepackage{blkarray, bigstrut}
\usepackage{amsthm}
\usepackage{algorithm}
\usepackage{algpseudocode}

\theoremstyle{definition}

\usepackage{mathtools}

\usepackage{array,amssymb,booktabs}
\newcolumntype{C}{>{$}c<{$}}
\AtBeginDocument{
    \heavyrulewidth=.08em
    \lightrulewidth=.05em
    \cmidrulewidth=.03em
    \belowrulesep=.65ex
    \belowbottomsep=0pt
    \aboverulesep=.4ex
    \abovetopsep=0pt
    \cmidrulesep=\doublerulesep
    \cmidrulekern=.5em
    \defaultaddspace=.5em
}

\usepackage[usenames, dvipsnames]{color}

\begin{document}

\preprint{APS/123-QED}

\title{Experimental evaluation of digitally-verifiable photonic computing for blockchain and cryptocurrency}

\author{Sunil Pai}
\email{sunilpai@stanford.edu}
\affiliation{Department of Electrical Engineering, Stanford University, Stanford, CA 94305, USA}
\author{Taewon Park}
\affiliation{Department of Electrical Engineering, Stanford University, Stanford, CA 94305, USA}
\author{Marshall Ball}
\affiliation{Courant Institute, New York University, NY, USA}
\author{Bogdan Penkovsky}
\affiliation{PoWx, Cambridge, MA, USA}
\author{Maziyar Milanizadeh}
\affiliation{Politecnico di Milano, Milan, Italy}
\author{Michael Dubrovsky}
\affiliation{PoWx, Cambridge, MA, USA}
\author{Nathnael Abebe}
\affiliation{Department of Electrical Engineering, Stanford University, Stanford, CA 94305, USA}
\author{Francesco Morichetti}
\affiliation{Politecnico di Milano, Milan, Italy}
\author{Andrea Melloni}
\affiliation{Politecnico di Milano, Milan, Italy}
\author{Shanhui Fan}
\affiliation{Department of Electrical Engineering, Stanford University, Stanford, CA 94305, USA}
\author{Olav Solgaard}
\affiliation{Department of Electrical Engineering, Stanford University, Stanford, CA 94305, USA}
\author{David A.B. Miller}
\affiliation{Department of Electrical Engineering, Stanford University, Stanford, CA 94305, USA}

\begin{abstract}
    As blockchain technology and cryptocurrency become increasingly mainstream, ever-increasing energy costs required to maintain the computational power running these decentralized platforms create a market for more energy-efficient hardware. Photonic cryptographic hash functions, which use photonic integrated circuits to accelerate computation, promise energy efficiency for verifying transactions and mining in a cryptonetwork. Like many analog computing approaches, however, current proposals for photonic cryptographic hash functions that promise similar security guarantees as Bitcoin are susceptible to systematic error, so multiple devices may not reach a consensus on computation despite high numerical precision (associated with low photodetector noise). In this paper, we theoretically and experimentally demonstrate that a more general family of robust discrete analog cryptographic hash functions, which we introduce as LightHash, leverages integer matrix-vector operations on photonic mesh networks of interferometers. The difficulty of LightHash can be adjusted to be sufficiently tolerant to systematic error (calibration error, loss error, coupling error, and phase error) and preserve inherent security guarantees present in the Bitcoin protocol. Finally, going beyond our proof-of-concept, we define a ``photonic advantage'' criterion and justify how recent developments in CMOS optoelectronics (including analog-digital conversion) provably achieve such advantage for robust and digitally-verifiable photonic computing and ultimately generate a new market for decentralized photonic technology.
\end{abstract}

\maketitle

Photonic integrated circuits consisting of networks or ``meshes'' of Mach-Zehnder interferometers (MZIs) \cite{Miller2013Self-configuringInvited, Bogaerts2020ProgrammableCircuits} are typically proposed as time- and energy-efficient matrix multiplication accelerators for analog domain applications such as quantum computing \cite{Arrazola2018MachineComputers, Carolan2015UniversalOptics}, sensing, telecommunications \cite{Annoni2017UnscramblingModes} and machine learning \cite{Shen2017DeepCircuits}. Since photonic meshes can be designed and mass-produced using well-established silicon foundry processes, there has recently been increased effort to commercialize the technology for analog domains that do not necessarily require high accuracy for high performance (e.g. machine learning). In this work, we extend applications of photonic meshes from the continuous analog domains of sensing and quantum computing to discrete and digital domains of cryptography and blockchain technology. The problem we want to solve is to design photonic matrix multiplication hardware successfully under more stringent numerical accuracy requirements that require nearly ``perfect'' digital computation. Essentially, we are exploring the use of such meshes as digital rather than analog multipliers. 

As our core application, we explore ``photonic blockchain'' technology which implements ``optical proof of work'' (oPoW) \cite{Dubrovsky2020TowardsWork}, proof that this discrete matrix multiply computational work has been performed in the optical domain. 
Equipped by optical computation with sufficient accuracy, such protocols can leverage energy efficient computation and state-of-the-art photonic hardware (such as photonic meshes) to verify cryptocurrency transactions and ultimately other wide ranging applications of blockchain such as medical data, smart contracts, voting, logistics and tracking, spam filters and protection from distributed denial-of-service (DDoS) attacks.
In situations where energy cost is a bottleneck, blockchain technologies that use optical proof of work inherently incentivize using photonic hardware over other alternatives to gain competitive advantages in compute efficiency and further security against malicious actors such as malware or attack vectors. 
For example, at the time of writing, cryptocurrency mining accounts for as much energy as all data centers globally and this energy consumption will increase, by design, as more value is stored in decentralized PoW blockchains.
Energy cost concerns have contributed to recent crashes in the cryptocurrency marketplace, including over the span of two weeks in May 2021 and again in Jan 2022 that reduced market capitalization of popular cryptocurrencies by the equivalent of more than 1 trillion dollars.
Photonic blockchain could thus serve as a timely application for a proof-of-work based cryptocurrency that incentivizes energy-efficient photonic hardware, which can furthermore prove to be an appealing option for other blockchain-based applications.

While the primary emphasis in this paper is on the photonic implementation and performance of optical proof of work, we must first clarify the energy-efficiency problem in blockchain and cryptocurrency. Cryptocurrency is a decentralized currency market where transactions (e.g. ``Alice gives Bob 1 bitcoin'') are stored in a chain of blocks (``blockchain''). To earn a share of the market, a cryptocurrency miner can ``mine" (add a new block of transactions) to the blockchain using computational ``proof of work'' (PoW) where the computer can solve a puzzle for a payout reward. This puzzle consists of generating a 256-element bitvector (vector of 1's and 0's) by feeding digital block transaction data through a cryptographic hash function $H$, such as SHA-256 (a map which converts any digitally encoded data into 256-bit numbers and which is infeasible to invert), which is the ``computational work.'' \footnote{Such hash functions, ``one-way'' (non-invertible) functions for private-key cryptography, are more generally used to securely encrypt and decrypt data for various secure applications beyond blockchain.} This function is called twice (once on the original block data and again on the result of the first call) through a scheme called  ``HashCash'' while adjusting a \textit{nonce} (32-bit pseudorandom number) in the block until the first $B$ bits in the bitvector are 0, which proves that sufficient computational work has been done and adds the block to the blockchain.
The parameter $B$ is a tunable difficulty parameter that is increased as the coin (which in many cases has limited supply) is more scarce and the expected number of cycles before the puzzle is solved is $2^B$. 
Crucially, as it relates to this paper, cryptocurrency mining comes at an energy cost proportional to the number of hash function solves before a block is mined and transactions in the block are considered verified. 

In optical PoW, the miner is incentivized to reduce mining costs by choosing optical (photonic) hardware that improves the energy efficiency and speed of computation compared to digital alternatives. 
However, the great challenge of such computing is that \textit{any} error in the bits output by the analog hardware renders an entire hash verification invalid; this necessitates some strict design criteria and possibly some error correction coding which we explore in this paper. 
We address this accuracy problem by numerically and experimentally evaluating a new photonic hash function called ``LightHash,'' a modified hash function from Bitcoin's Hashcash that combines the energy-efficiency of linear optics with the security assurances of the Bitcoin protocol. 
We define feasible design criteria (e.g., number of photonic inputs and outputs) and propose a hardware agnostic error correction scheme that enables our photonic hash function to outperform any digital alternative.

\section*{Photonic blockchain}

Photonic blockchain can be defined as a modification of the Bitcoin protocol hash functions using what we refer to as a ``photonic hash function,'' which is a type of optical cryptography.
In optical cryptography, a cryptographer encrypts or decrypts a message by sending the message bits through a hash function, where at least one part of the hash function requires an optical device.
Here, we explore photonic hash functions that incorporate a ``universal'' photonic mesh \cite{Dubrovsky2020TowardsWork} or $N$-port triangular or rectangular MZI networks  \cite{Miller2013Self-configuringInvited, Reck1994ExperimentalOperator, Clements2016AnInterferometers}. 
Meshes operate by repeatedly interfering spatially multiplexed mode vectors of coherent light (over the $N$ ports), where modes are represented as complex numbers with amplitude and phase. 
The constructive and destructive interference can be programmed using electrically-controlled phase shifts to implement any unitary transmission matrix $U \in \mathrm{U}(N)$ (satisfying the energy conserving property $U^\dagger U = I$) \cite{Miller2013Self-configuringInvited, Reck1994ExperimentalOperator}.
After adding a column of ``singular value'' MZIs followed by a second universal network as in Ref. \cite{Miller2013Self-configuringInvited}, it is possible to compute an arbitrary linear operator based on the singular value decomposition (SVD) of any matrix $Q$ \cite{Miller2013Self-configuringInvited}.
The resulting photonic processors can be programmed to implement arbitrary energy-efficient linear operations; though energy must be spent in generating, modulating, and detecting the optical signals, the actual matrix-vector product is performed by passive linear optical transformations without additional power.
The rest of the computation in photonic hash functions include logical operations on bits that are best implemented in the digital domain (e.g. SHA-256 is efficiently implemented on digital processors) and ultimately provide the necessary provably secure protection.
By co-integrating this digital functionality with photonic meshes in a systematic manner, we leverage the unique benefits of optics (linear computation) and electronics (nonlinear computation and logic) for a fully integrated photonic cryptographic solution.
While previous proposals of this scheme exist (e.g. HeavyHash \cite{Dubrovsky2020TowardsWork}), a protocol that is sufficiently error tolerant and is both time-and energy-efficient (including the analog-digital conversion) is yet to be proposed.
Similar challenges are faced in photonic circuits for digital optical telecommunications, and indeed, the mathematics of ``bit error rates'' also can be applied to the problem of optical cryptography.
Ultimately, the core challenge is to find a protocol that successfully brings photonic computing, a technology typically used for analog computing, into the digital realm with near perfect accuracy.

To this end, we examine whether meshes can accurately implement matrix multiplication compared to an electronic digital implementation so they could ultimately be used for PoW cryptography and confer a ``photonic advantage.''
The ``photonic advantage'' for our particular scheme follows from the conjecture that, within the LightHash evaluation, photonic hardware performs amortized matrix multiplication by a random block-diagonal $Q$ operator at least an order of magnitude more efficiently than traditional hardware, where element within the blocks of $Q$ is sampled from uniform distributions over a set of $K$ integers.
First, through numerical simulation, we show that programming a block integer matrix $Q$ onto a series of SVD-based photonic architectures \cite{Miller2013Self-configuringInvited} (as opposed to purely unitary circuits \cite{Reck1994ExperimentalOperator}) and adjusting the numerical precision through different integer $K$ values can minimize the systematic error in the analog computation to make it more amenable to optical cryptography.
Then, we experimentally evaluate the cryptographic protocol on a physical photonic chip accelerator capable of performing $4 \times 4$ unitary matrix-vector products to estimate performance on our new proposed LightHash protocol.
Since the LightHash matrix-vector operation is performed in discrete space, we can find conditions such that possible outputs are separated sufficiently far enough to guarantee near-perfect accuracy.
The increased energy-efficiency-per-compute of a photonic platform would not only help to increase the security of cryptocurrencies and blockchain operations generally but also would result in a significant shift from energy cost (operating expense) to resource cost (capital expense) in cryptocurrency mining \cite{Dubrovsky2020TowardsWork}.
The resulting increased demand for photonic chips could incentivize photonic integrated circuit (PIC) development and manufacturing by adding new applications.

\begin{figure*}
    \centering
    \includegraphics[width=\textwidth]{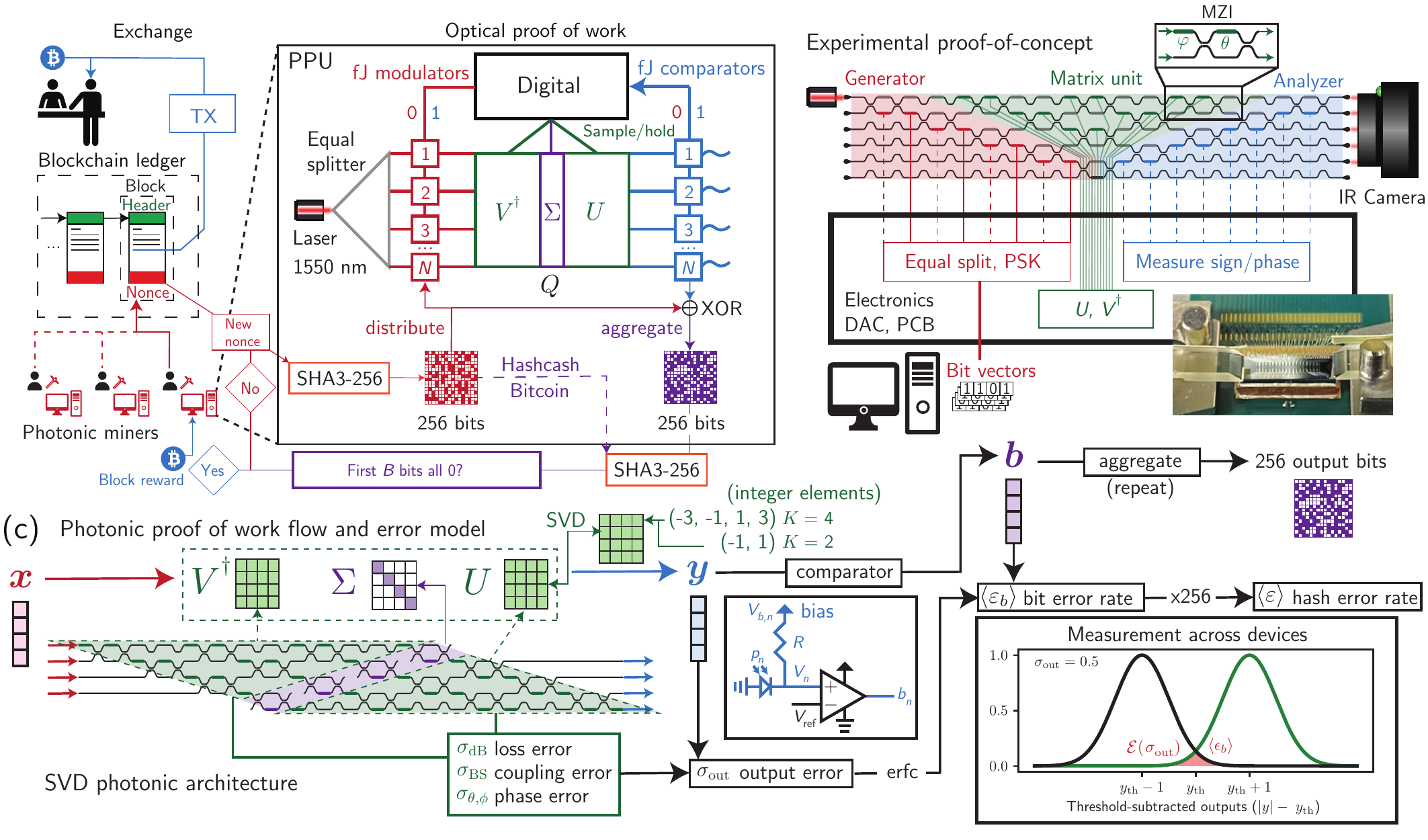}
    \caption{(a) The LightHash optical proof of work protocol, similar to HeavyHash \cite{Dubrovsky2020TowardsWork}, is a slight modification of the Bitcoin protocol, where an arbitrary photonic mesh-based matrix-vector product is inserted in the middle.  Transactions are verified by photonic miners as in the Bitcoin network, with photonic chips being the ideal technology to achieve a block reward. Output bits are measured using femtojoule comparators ideally running at GHz speeds. (b) The experimental setup used to evaluate the LightHash protocol ($N = 4$, variable $K$) involves running $U, V$ on-chip, responsible for most of the error, and multiplying singular values off-chip. (c) The photonic proof of work error analysis model shows how systematic error in the device mathematically propagates all the way to an overall hash error rate which arises due to overlap between successive values near the threshold shown in the inset. Reducing the hash error rate is the main aim of this work and is necessary to implement LightHash in practice.}
    \label{fig:opow}
\end{figure*}

\section*{Algorithm}

As shown in Fig. \ref{fig:opow}, our photonic cryptocurrency protocol incorporates a photonic processor unit (photonic mesh) within Bitcoin's proof-of-work hash computation to implement LightHash.

Our LightHash photonic cryptographic protocol transforms block (transaction) data into a 256-bit ``possible solution'' to a cryptographic puzzle, and includes a photonic integrated chip computation within the protocol. The protocol begins with the well-known SHA3-256 protocol, which is part of the already-prevalent digital cryptocurrency Bitcoin and converts block data (containing transactions in the marketplace) into a 256-bit vector containing a sequence of 256 0's and 1's. This bit data is directly fed to optical modulators that control the optical input into the photonic accelerator chip in chunks of $N$ bits, which has the following behavior:

\begin{enumerate}
    \item \textbf{Input}: the input into the photonic network is a phase-shift keyed bitstream $\boldsymbol{b}_{\mathrm{in}}$ which is represented as inputs of equal magnitude set to either $x_n = \{1, -1\}$ depending on a bit value $b_n = \{0, 1\}$ so, a vector of inputs $\boldsymbol{x} = e^{i\pi\boldsymbol{b}_{\mathrm{in}}} / \sqrt{N}$; this can be set by sending digital signals to well-calibrated optical modulators.
    \item \textbf{Device operator}: As shown in Fig. \ref{fig:opow}(a, c), the device operator for each block $Q_m = U\Sigma V^\dagger$ with circuit size $N$ consists of two unitary operators $U, V$ of size $N$, implemented using triangular or rectangular networks \cite{Reck1994ExperimentalOperator, Clements2016AnInterferometers} and a set of $N$ singular values $\Sigma$, implemented using MZI node attenuators (with ``drop ports'') \cite{Miller2013Self-configuringInvited}. The elements $Q_{m, ij}$ are randomly sampled to be one of $K$ distinct integers centered symmetrically around $0$ and spaced $2$ apart. At block creation only, a digital computer is used to find the static phase shifts for meshes implementing $U, V, \Sigma$ to ultimately program $Q$ onto the chip.
    \item \textbf{Output}: The output of the device is the complex output vector $\boldsymbol{y} = Q \boldsymbol{x}$ with output power $\boldsymbol{p} = |\boldsymbol{y}|^2$, where $|\cdot|^2$ is an elementwise absolute value-squared operation. A photodetector equipped with a transimpedance element (load resistor or amplifier) converts power to voltage, which is then fed through output comparators corresponding to threshold power $p_{\mathrm{th}} = y_{\mathrm{th}}^2$ to determine output bits $\boldsymbol{b} := H(\boldsymbol{p} - p_{\mathrm{th}})$ (where $H$ is the Heaviside step function). At block creation, selection of $p_{\mathrm{th}}$ via simulation guarantees roughly equal probability of a 0 or 1 output bit.
\end{enumerate}

Note that the definition of the threshold amplitude $|y_{\mathrm{th}}|$ should be consistent with the scaling of the blocks in matrix $Q$. Since the maximum singular value of $Q$ is set to 1 in the physical implementation (no optical gain elements are used in our photonic mesh), the threshold amplitude is also scaled by this factor as discussed in more detail in the Appendix.

A unique feature of LightHash is the numerical resolution $K$, which can be used to change the range of possible output values. For instance, $K = 2$ means the matrix elements can be either $1$ or $-1$, and $K = 4$ means the options for each matrix element are $(-3, -1, 1, 3)$. Each row vector-vector product in the overall matrix vector product can actually be thought of as a random walk with $K$ defining possible step sizes ($1$ for $K = 2$ and $(1, 3)$ for $K = 4$). Since the inputs are either $-1$ or $1$, an increase in $K$ means an increase in range of possible output values and effectively the number of bits or quantized levels present in the output. Due to a larger required number of output bits, the use of higher $K$, as with higher $N$, leads to higher computational efficiency but a more error-prone photonic chip.

Note that the device is set to implement $Q$ only once per block added to the blockchain, which means that the photonic miner has some time to self-configure itself to implement $Q$, and block times can generally be several minutes at sufficiently high difficulty. If $N < 256$, we repeat $256 / N$ times (assuming $N$ divides 256) to output a total of 256 bits that is ``exclusive or'd'' (XOR'd) with the original input vector and fed into the second SHA3-256 function as in Fig. \ref{fig:opow}(a). Additional context for the design choices here is provided in the Appendix; the gist is that our formulation makes optical proof of work feasible in the presence of error.

Aside from random laser and photodetector noise, there are three types of non-random errors that are more challenging to address and on which we focus on in this paper: phase error, coupling error, and loss error. These sources of error can be compensated in various ways using phase shifter calibration in a photonic mesh; for example, self-configuration  \cite{Miller2013Self-configuringInvited, Hamerly2021AccurateInterferometers} or off-chip calculation \cite{Bandyopadhyay2021HardwarePhotonics} can to some extent compensate for these errors. (Note that while phase and coupling error can be mostly compensated using error correction or self-configuration, error due to unequal losses cannot be compensated in this way.) As discussed in the Appendix, calibration and error correction generally occurs at the level of individual ``unit cell'' nodes or Mach-Zehnder interferometers (MZIs) that can be more straightforwardly characterized.

To correct this error, we use a simple form of ``hardware-agnostic error correction'' in which the computation is repeated up to $R$ times across $R$ circuit copies. If the expected error is $\sigma_{\mathrm{out}}$, this can reduce the error to $\sigma_{\mathrm{out}} / \sqrt{R}$, a factor of $\sqrt{R}$ improvement. This repetition may be implemented using $R$ separate devices implemented on the same chip. To save on energy consumption, we can use the same number of modulators and split the input signal $\boldsymbol{x}$ across $R$ different meshes implementing the same $Q$ but different error, in a process called ``hardware agnostic error correction'' as each mesh samples a presumably random systematic error. One potential drawback of this approach is that the systematic error may be correlated across the $R$ meshes, so one way to ensure uncorrelated error is to permute the singular values (and appropriate basis vectors of $U, V^\dagger$). The number of comparators might also stay the same assuming the photocurrents from corresponding photodetectors at the $R$ device operator outputs can be grouped into a single current, and then passed through a transimpedance amplifier and comparator.

The challenge of photonic cryptography that we address is to ensure accuracy across all 256 bits (hash or packet error rate, PER) while also affording a significant advantage over equivalent digital hardware in speed and energy efficiency (otherwise there would be less demand for photonic hardware). For simplicity, we may consider all bits to have independent bit error rates (BER) $\varepsilon_b$ so the hash error rate is given by $\varepsilon = 1 - (1 - \varepsilon)^{256} \approx 256 \varepsilon_b$. To put this in practical terms, for any given device to have $1\%$ PER, each of the individual bits should have roughly $0.004\%$ BER. This increases the importance of error correction in photonic integrated circuits, which is especially challenging in the presence of unbalanced photonic loss. This requires exploring the tradeoffs of increasing circuit size $N$ and difficulty $K$ (which increase the difficulty) over the error. Pseudocode for the entire optical proof of work based on LightHash is provided in the Appendix.

\begin{figure*}
    \centering
    \includegraphics[width=\textwidth]{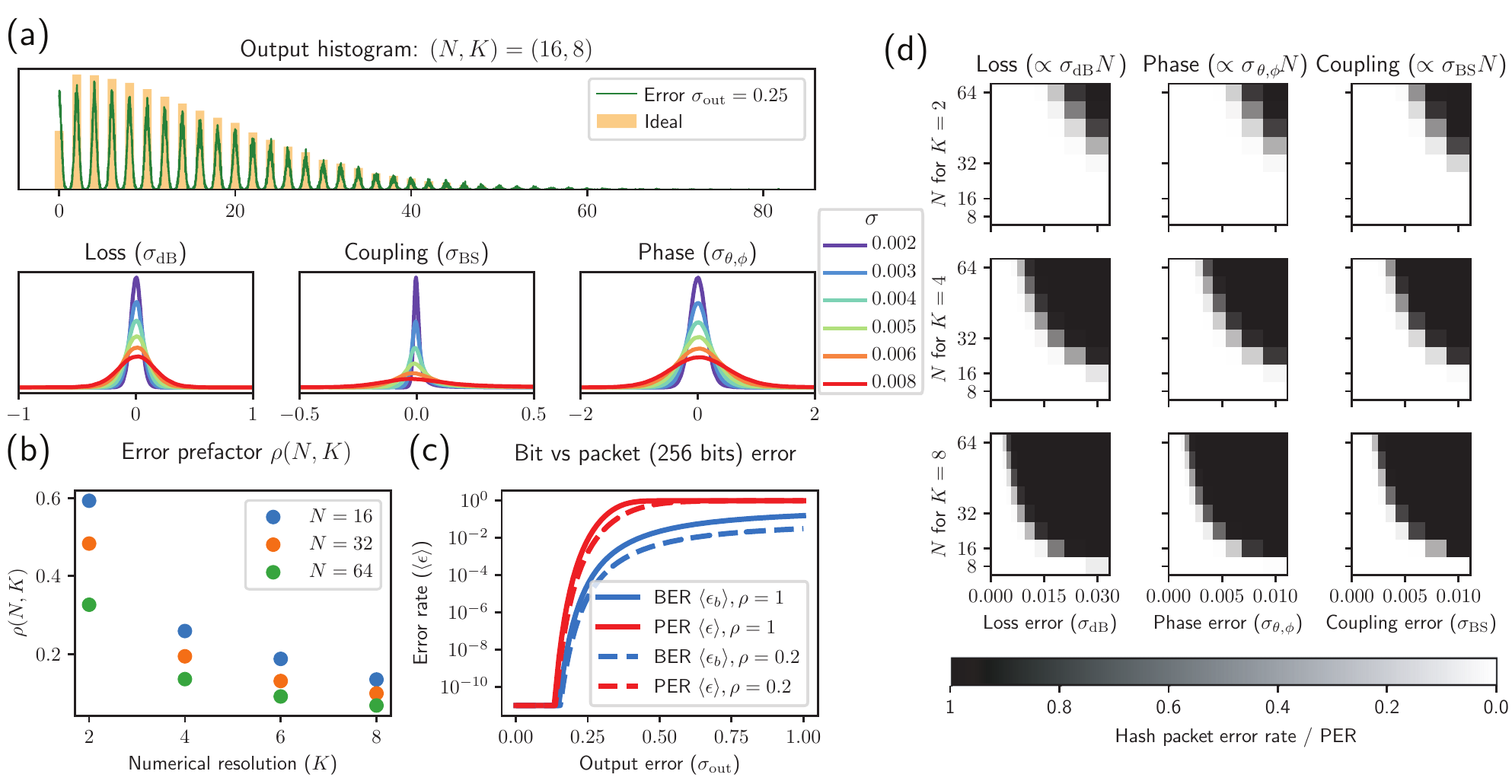}
    \caption{(a) Example output histogram for $N = 16$ and $K = 8$ exceeds the capabilities of our device ($N = 4$) shown for both ideal and experimental implementations. Overall coupling error, loss error and phase error contributions for various component-wise $\sigma < 0.01$ all are roughly Gaussian with coupling errors skewed slightly to the right. (b) Error scaling $\rho(N, K)$ can be empirically determined found via simulation to determine the error at threshold $p_{\mathrm{th}}$. (c) The error prefactor $\rho$ has a small effect on the bit error scaling; regardless of the scaling of $\rho$, we find that $\sigma_{\mathrm{out}} = 0.25$ is sufficient to ensure sufficiently low bit error ($<1\%$) throughout the entire range of $\rho$ in (b). (d) Bit threshold error profile shows sharp transition in overall hash error as a function of circuit size $N$ and/or loss, phase, and coupling errors that increases with $K = 2, 4, 8$ (1 to 3 bits), establishing performance bounds for each error type.}
    \label{fig:simulation}
\end{figure*}

\section*{Scaling simulations}

First, we numerically evaluate increasing circuit sizes $N = 8$ to $64$ for $K = 2, 4, 8$.

Ideally, the outputs $\boldsymbol{y}$ follow a roughly discretized Gaussian distribution as might be expected by a random walk based on our definitions of $Q$ and $\boldsymbol{x}$ (The matrix vector multiplication implements exactly a random walk of $N$ time steps for $K = 2$.). The field magnitudes $|\boldsymbol{y}|$ are more readily measured by output photodetectors and as expected form a discrete half-normal distribution as shown in Fig. \ref{fig:simulation}(a), with a notable dip in histogram values for outcomes equal to zero.

The simulation outputs are measured over a large range of possible $Q, \boldsymbol{x}$, and we assume that this is equivalent to simulating many devices implementing the same operator over many inputs. The resulting histogram of outputs $|\boldsymbol{y}|$ form error distributions with roughly similar standard deviations $\sigma_{\mathrm{out}}$. Evaluated over MZIs across the photonic mesh, the loss standard deviation error $\sigma_{\mathrm{dB}}$ is in dB, the phase standard deviation error $\sigma_{\mathrm{\theta, \phi}}$ is in radians, and the coupling standard deviation error $\sigma_{\mathrm{BS}}$ is also in radians scaled by 50/50 coupler beat length. Further details on these error source models are discussed in the Appendix.

These different types of errors all increase linearly with $K$ and $N$ and thus add in quadrature to approximately give the overall error:
\begin{equation} \label{eqn:outputerror}
    \sigma_{\mathrm{out}}^2 \approx N^2K^2 (k_{\mathrm{\theta, \phi}} \sigma_{\mathrm{\theta, \phi}}^2 + k_{\mathrm{BS}} \sigma_{\mathrm{BS}}^2 + k_{\mathrm{dB}} \sigma_{\mathrm{dB}}^2),
\end{equation}
which we generally observe in Fig. \ref{fig:simulation}(a) assuming a minimal thermal or shot noise contribution. (This minimal noise contribution is a valid assumption for our photonic mesh as shown in the Appendix and in general, can be compensated by integrating longer or changing photodetector circuitry.) As is evident in Fig. \ref{fig:simulation}(a), error distributions are roughly Gaussian, though our simulations suggest that the coupling error $\sigma_{\mathrm{BS}}$ results in a long ``tail'' compared to the loss and phase error types which are more Gaussian. One possible reason for this may have to do with the nonlinear transformation of phase errors equivalent to these coupling errors \cite{Bandyopadhyay2021HardwarePhotonics, Hamerly2021InfinitelyInterferometers}.

As previously shown in Fig. \ref{fig:opow}(c), bit errors arise when there is ``bit threshold overlap'' in the (approximately) Gaussian error distributions between successive values around the threshold, given by $\mathcal{E}(\sigma_{\mathrm{out}}) = 0.5\mathrm{erfc}((\sigma_{\mathrm{out}} \sqrt{2})^{-1})$, where $\mathrm{erfc}$ denotes the error function. To find the corresponding expected bit error, we multiply the overlap in error by twice the probability $\rho(N, K)$ (the densities in the histogram of Fig. \ref{fig:simulation}(a)) that the values belong to the Gaussian spikes immediately before or after the threshold $y_{\mathrm{th}}$. Assuming $\langle \varepsilon_b \rangle$ is small, we get the expression for expected hash error $\langle \varepsilon \rangle$:
\begin{equation}
    \langle \varepsilon \rangle \approx 256\rho(N, K)\mathcal{E}(\sigma_{\mathrm{out}}).
\end{equation}
The analysis of scaling of the bit error rate is shown in Fig. \ref{fig:simulation}(b, c) to indicate relatively small contribution of the prefactor for output errors except in the transition region between high and low error around $\sigma_{\mathrm{out}} \approx 0.25$.

The main finding from our simulations in Fig. \ref{fig:simulation}(d) is that high-fidelity (feasible) operation for appreciably large $N$ generally requires error standard deviation much less than 0.01 radians (coupling, phase) or 0.03 dB (loss) which can be considered to be empirical error bound requirements. Such a design at the component level is challenging and depends heavily on the implementation (calibration) and fabrication of the device. While calibration and self-configuration may sidestep such issues in sufficiently small circuits \cite{Miller2013Self-configuringInvited, Annoni2017UnscramblingModes}, such stringent requirements are worth noting when designing devices for cryptographic applications.

\section*{Experimental evaluation}

\begin{figure*}
    \centering
    \includegraphics[width=\textwidth]{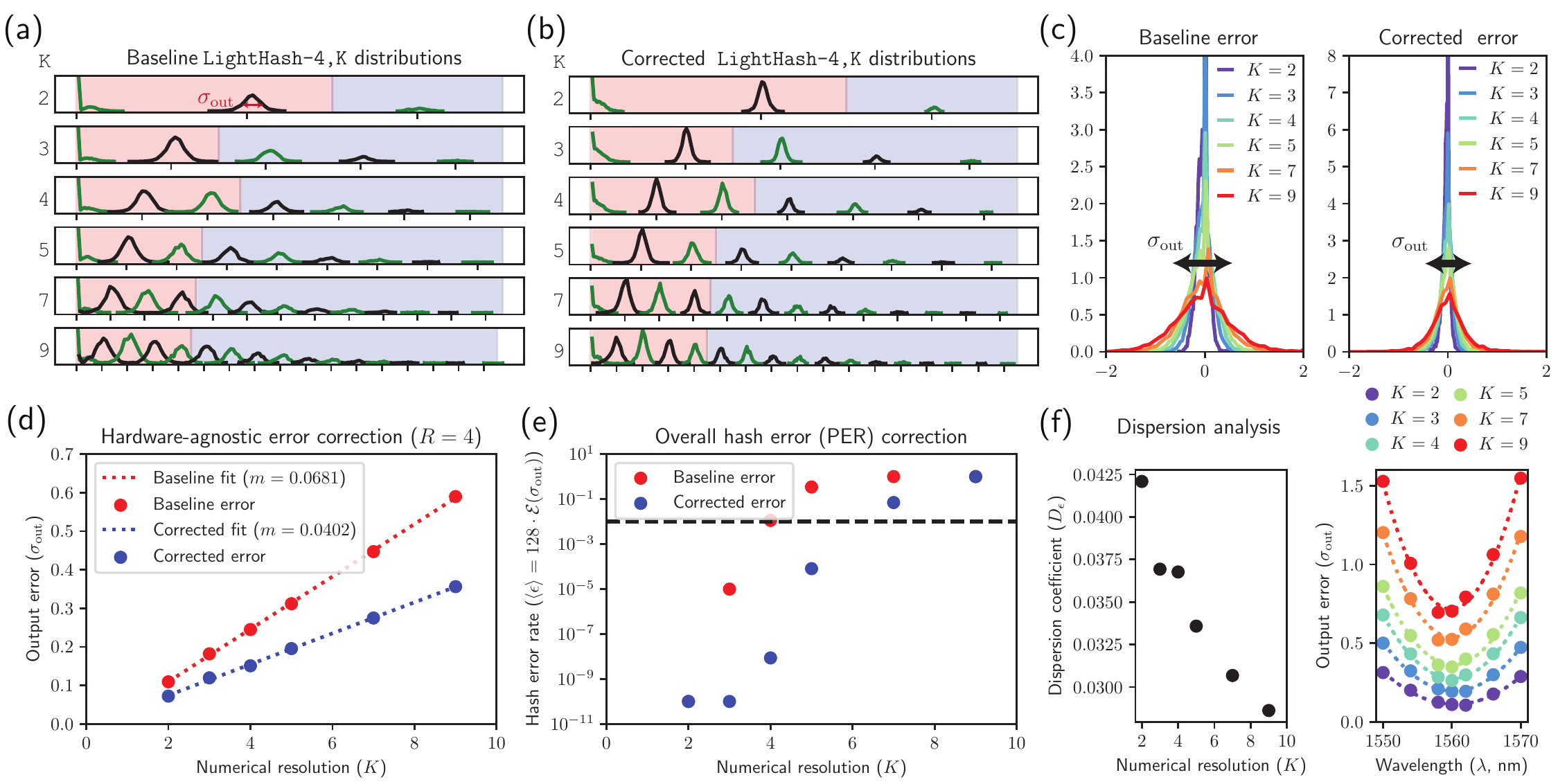}
    \caption{Outcome LightHash histograms for 250 random matrices $Q$ given $N = 4$ for varying $K$ for (a) baseline and (b) hardware-agnostic error-corrected implementations. We label alternating colors green and black to clearly delineate the overlap regions between successive values (spaced 2 apart) labelled by tick marks. The red and blue regions correspond to a bit assignment of 0 and 1 respectively by digital thresholding. (c) Comparison between the baseline and hardware agnostic error correction error distributions (subtracting the ideal values from the outcome histograms in (a) and (b)). (d) The standard deviation of the error, $\sigma_{\mathrm{out}}$, is roughly proportional to $K$ for both the baseline and corrected implementations, with the corrected implementation having a much smaller slope $m$. (e) Sharp transition in feasibility is demonstrated for the baseline and error-corrected cases as a function of $K$ (similar behavior is expected for $N$) (f) Dispersion of the error given calibration at the center wavelength 1560 nm shows parabolic increase in error around the center wavelength as expected, but the dispersion coefficient interestingly decreases with $K$.}
    \label{fig:experiment}
\end{figure*}

Now that we have defined our protocol and simulated the scalability of the technique, we experimentally quantify errors in a $4 \times 4$ port MZI mesh network (i.e., $N = 4$) as a function of numerical resolution $K$ using our custom designed chip and the experimental setup in Fig. \ref{fig:opow}(b). To estimate these errors, we record a distribution of output magnitudes at the network output given random $\boldsymbol{x}, Q$. Using this, we assume we can achieve an experimental estimate of $\sigma_{\mathrm{out}}$ measured across many devices. As expected and shown in Fig. \ref{fig:experiment}(a, b), the distribution follows a discretized half-normal distribution with Gaussian-distributed spikes at each of the possible outputs.

Next, as shown in Fig. \ref{fig:experiment}(c, d, e), we perform an error correction analysis by singular value permutation as previously proposed. The singular value decomposition is invariant given any permutation identically applied to the rows of $V^\dagger$, the columns of $U$, and the singular values of $\Sigma$, i.e., $Q = U \Sigma V^\dagger = (UP)(P\Sigma)(PV^\dagger)$, where $P$ is a matrix that implements the permutation. Therefore, error correction is possible by applying different $P$ to the $R$ meshes implementing $Q$. The proof of invariance is that $Q_{ij} = \sum_{k = 1}^{N} U_{ik}\Sigma_{kk} V^\dagger_{kj}$ and $k$ can be relabelled in any order, resulting in the same $Q$ by reflective property of addition. In our case, we average the result over four cyclic permutations of the singular values, i.e. $(1, 2, 3, 4), (4, 1, 2, 3), (3, 4, 1, 2), (2, 3, 4, 1)$. As expected, we find that the error gets roughly cut in half when $R = 4$ (the slope $m$ is reduced by about $41\%$). This suggests that averaging results over devices implementing $Q$ with permuted singular values can significantly reduce the error at the expense of increased device footprint.

Finally in Fig. \ref{fig:experiment}(f), we consider the ``error dispersion'' relation $\varepsilon(\lambda)$, exploring the effect of wavelength on the error to explore the possibility of parallelizing the computation over multiple wavelengths in a 20 nm wide band at our empirically determined optimal wavelength $\lambda_c = 1560$ nm:
\begin{equation} \label{eqn:dispersion}
    \varepsilon(\lambda) \approx \langle \varepsilon \rangle [1 + D_\varepsilon (\Delta \lambda) ^2],
\end{equation}
where $\Delta \lambda = \lambda - \lambda_c$ and $D_\varepsilon$ is the ``relative'' error dispersion (that depends on $N, K$) evaluated at $\lambda_c$. Our results in Fig. \ref{fig:experiment}(f) indicate that this relative dispersion coefficient as defined in Eq. \ref{eqn:dispersion} actually \textit{decreases} slightly with $K$. Note that there is an increase in absolute dispersion, but a decrease in the relative error dispersion.

\section*{Discussion and outlook}

Our results suggest that a digitally verifiable photonic mesh for proof-of-work applications such as cryptocurrency requires stringent design and fabrication requirements. In particular, for a large ensemble of cryptographic devices to agree on some computation, the systematic (biased) error rate has to be sufficiently small such that proof of work can be achieved with high probability. Our results indicate that past proposals of photonic cryptocurrency such as HeavyHash face hurdles to achieve such a success level. In this paper, we have taken a two-tier approach to improve on past proposals and ultimately address these challenges.

First, we propose the LightHash protocol which allows us to modulate the ``difficulty'' of a problem by changing the number of possible values (numerical resolution) $K$. For larger values of $N$ and $K$, we find that the likelihood of error is dramatically increased and generally find an approximate scaling for the output error $\sigma_{\mathrm{out}} \propto N K \sigma$, when $\sigma$ is a component-wise phase or coupling error (in radians) or component-wise loss error (in dB) localized to the phase shifters. This is an important design criterion that establishes fundamental limits on how high $N$ and $K$ can go before there is too much hash error $\langle \varepsilon \rangle$ for a blockchain or cryptocurrency scheme to be ``feasible.''

Second, we propose a new method of ``hardware agnostic error correction,'' which can help to reduce the error in a manner orthogonal to current error correction protocols such as self-configuration \cite{Miller2013Self-configuringInvited, Hamerly2021AccurateInterferometers}, hardware-aware error correction \cite{Bandyopadhyay2021HardwarePhotonics}, and gradient-based approaches \cite{Pai2019MatrixDevices}. By ``orthogonal,'' we refer to the fact that error correction can be applied using any of the currently known approaches to individual photonic meshes, and hardware agnostic error correction can be used to further reduce the error once those options are completely exhausted. 

Our results indicate that of the many routes achieve a feasible blockchain technology for cryptocurrency mining, sufficient reduction of systematic error $\sigma_{\mathrm{out}}$ will have the most effect on reducing hash error rate $\langle \varepsilon \rangle$. For instance, error correction resulting in a decrease of $\sigma_{\mathrm{out}}$ from 0.5 to 0.25 (using $R = 4$, which roughly multiplies device footprint by 4) can reduce $\langle \varepsilon_b \rangle$ by four orders of magnitude due to the exponentially-decaying tail of the Gaussian and ultimately greater than 99\% accuracy. In comparison, the choice of threshold $p_\mathrm{th}$ affects only the prefactor $\rho(N, K)$ in the ranges $N \in [8, 64], K \in [2, 9]$ and therefore has significantly less impact on the expected error $\langle \varepsilon \rangle$. This observation, in addition to Figs. \ref{fig:simulation}(c) and \ref{fig:experiment}(e), suggests that the barrier between feasibility and infeasibility is rather sharp and can be addressed by error correction to reduce $\sigma_{\mathrm{out}}$ mostly in cases where feasibility is marginal. This sharp boundary is explained by the fact that there is a quadratic exponent in the $\mathrm{erfc}$ function's integrand, i.e., $\mathrm{erfc}(z) \propto \int _{z}^{\infty}e^{-t^{2}} dt$.

Based on the results of this paper, there are several reasons to prefer a photonic blockchain and optical proof of work over a digital alternative to carry out the LightHash proof of work scheme. First, photonic cryptocurrency miners could scale to being more profitable since the energy efficiency and reduced latency lead to higher profits. With increased adoption, ``mining pools'' that use photonic hardware can lead to a frequent and consistent stream of income for a photonic miner compared to a conventional digital miner (Appendix). The reason for this improved efficiency is that most of the energy used by the LightHash protocol is nominally concentrated in output comparators that require at least $40$ fJ/bit \cite{Filippini2018AConversion, Miyahara2008AADCs}, assuming the chip operates at GHz speeds. The total energy for $N = 64$ LightHash becomes roughly $10.5$ pJ of energy which is two orders of magnitude less energy than the most energy efficient digital hardware for matrix-vector multiplication \cite{Nahmias2020PhotonicNetworks}. Further discussion is left to the Appendix. 

Second, photonic hardware used in hash protocols can also be used for other applications, i.e., the hardware is not necessarily an application specific device. Importantly, the very chip we use in this paper to explore cryptographic hash functions can also be used to perform inference tasks and backpropagation training in photonic neural networks \cite{Pai2022ExperimentallyNetworks}. This would suggest that photonic mining hardware has key advantages over pure digital application-specific hardware that implements energy-efficient cryptography but serve no other purpose.

\section*{Conflicts of Interest}
SP, MD, MB, and BP all own a nominal amount of optical bitcoin (oBTC), a cryptocurrency launched in 2021 that uses the HeavyHash (not LightHash) protocol for PoW and are involved with PoWx, a nonprofit organization dedicated to energy-efficient optical computing for cryptography. MB and BP also own a nominal amount of Kaspa. SP, MD, BP, MB, SF, OS, DM are coinventors on a provisional patent application for LightHash, Prov. Appl. No.: 63/323727. DM holds patents on the SVD photonic mesh architecture. The authors declare no other conflicts of interest.

\section*{Contributions}
SP taped out the photonic integrated circuit and ran all experiments with input from TP, NA, MM, FM, AM, OS, SF, DM. SP conceptualized and simulated the LightHash protocol with input from BP, MD and MB. TP designed the custom PCB with input from SP. SP wrote the manuscript with input from all coauthors.

\section*{Data and software}
All software and data for running the simulations and experiments are available through Zenodo \cite{Pai2022Solgaardlab/photoniccrypto:Cryptocurrency} and Github through the Phox framework, including our experimental code via Phox \cite{Pai2022Phox:Devices}, simulation code via Simphox \cite{Pai2022Simphox:Library}, and automated photonic circuit design code via Dphox \cite{Pai2022Dphox:Design}.

\section*{Acknowledgements}
We would like to acknowledge help from Advanced MicroFoundries (AMF) in Singapore and for their help in building the necessary circuitry for our demonstration. We would also like to acknowledge funding from Air Force Office of Scientific Research (AFOSR) through grants FA9550-17-1-0002 in collaboration with UT Austin and FA9550-18-1-0186 through which we share a close collaboration with UC Davis under Dr. Ben Yoo. Thanks also to Payton Broaddus for helping with wafer dicing, Nagaraja Pai for advice on electronics and thermal control packaging, and Carsten Langrock and Karel Urbanek for their help in designing our movable optical breadboard. Finally we would like to thank Dirk Englund, Ryan Hamerly, and Saumil Bandyopadhyay for fruitful discussions on error modelling as well as Mustafa Hammood, Vsevolod Hulchuk, and Maxim Karpov for illuminating feedback.

\section*{Methods}

\subsection{Chip design}

We have designed a $6 \times 6$ photonic mesh chip fabricated by Advanced Micro Foundries (AMF) in their silicon-on-insulator platform capable of implementing $4 \times 4$ matrix-vector multiplication. The phase shifters controlling the generator and the mesh itself are all titanium nitride (TiN) and are all calibrated to achieve up to $2 \pi$ phase shift as a polynomial function of the square-voltage applied across each of the phase shifters. The calibration proceeds by sending light progressively to each MZI in the device starting from the left-most to the right-most MZI and sweeping phase shifts from $0$ to $5$ volts. More details on this calibration are provided in the Supplemental Material. One feature of our mesh design is that there are grating taps at each of the waveguide segments of the MZIs capable of outputting a small fraction ($\sim 3\%$) of the power in the guide. These are used to measure the powers after \textit{each} MZI to calibrate phase shifts and are also used to measure the outputs of the device using an IR camera.

\subsection{Experimental setup}

The photonic mesh chip is wirebonded by Silitronics Solutions to our custom-built PCB designed to interface with an NI PCIe-6739 controller for setting programmable phase shifts throughout the device. The input optical source is a Agilent 81606A tunable laser with a tunable range of 1460 nm to 1580 nm. To measure powers coming out of grating taps, we use an infrared (IR) Xenics Bobcat 640 series camera set to ``raw'' mode connected to an IR/visible microscope with an infinity corrected Mitutuyo 10X IR objective lens attached to a 40 cm tube lens. This optical setup is fixed to a movable Applied Scientific Imaging (ASI) stage to image optical powers emitted from the grating taps. 

Automation of the LightHash algorithm is accomplished via USB/GPIB connections to the tunable laser, MXIe-PCIe slot connection to the NI control board, and ethernet connection to the camera for measurements and calibration. A graphical user interface is designed using Holoviews/Bokeh to debug the device and analyze the calibration. The chip consists of $6 \times 6$ photonic meshes with grating inputs and fiber array optical interconnects (constructed by W2 Optronics) shown in the image of Fig. \ref{fig:opow}(b). These are wirebonded to a custom PCB ordered via Advanced Circuits which also features a thermistor and thermal connection to a thermoelectric cooler on an aluminum mount for efficient thermal stabilization based on a feedback loop.

\subsection{Calibration and operation}

In order to operate the mesh as a $4 \times 4$ matrix-vector multiplier, we couple the tunable coherent laser to the top input of the device via the fiber array interconnect and use the first diagonal ``row'' of MZIs to function as an optical setup machine or generator \cite{Miller2017SettingMethod, Miller2020AnalyzingNetworks}. Full details and associated figures for our calibration follow the same procedure as in \cite{Pai2022ExperimentallyNetworks}, but we provide a summary here.

We take an initial reference image to get a background and then to measure the spots intensities or powers, we sum up the reference-subtracted pixel values that ``fill'' the appropriate grating taps throughout the device. We ensure that saturation does not take place by reducing the laser power to approximately 50 $\mu$W at 20 ms integration time, which suggests nW camera pixel sensitivity as expected. The output powers are all normalized based on the total power in the system (sum of all grating taps along a column of MZIs or waveguides), which automatically removes any laser power fluctuation not originating from the photodetector measurement (i.e. from the laser source itself). The units of power used in this paper are based on renormalizing this power based on the input into the system such that the total power propagating through a column of waveguides is 1.

Because our architecture is capable of implementing only unitary matrices (not an SVD architecture), we elect to perform the singular value operation ($\Sigma$) on the computer and the majority of the computation (unitary matrices $U, V^\dagger$) on the computer. Since only four phase shifters are required to implement these singular value operations (versus 32 for the unitary operators), we assume that the experimental evaluation of the overall SVD architecture is roughly the same as multiplying by the appropriate singular values and evaluating $U, V^\dagger$ separately as indicated in the green box of Fig. \ref{fig:opow}(b).

\subsection{Node error model}
As discussed in the main text, we explore three sources of error in simulation: loss, coupling and phase. Each source of error arises from various fabrication imperfections or phase drift sources. In order to formalize these error contributions, we define an ideal MZI node and contrast the ideal operation from the non-ideal operation (phase, coupling, and loss error). All of our calculations are performed using our open source Python photonic simulation code simphox \cite{Pai2022Simphox:Library}.

The ideal MZI node in terms of building blocks that consist of a $\phi$ phase shift, 50:50 coupler, $\theta$ phase shift, and another 50:50 coupler, giving us the following mathematical representation acting on modes $x_1, x_2$ and yielding outputs $y_1, y_2$:
\begin{equation} \label{eqn:mzi}
    \begin{aligned}
    \begin{bmatrix} y_1 \\ y_2 \end{bmatrix} &= i\begin{bmatrix}e^{i\phi}\sin \frac{\theta}{2} & \cos \frac{\theta}{2} \\ e^{i\phi}\cos \frac{\theta}{2} & -\sin \frac{\theta}{2}
    \end{bmatrix} \begin{bmatrix}x_1 \\ x_2 \end{bmatrix}\\
    \boldsymbol{y} &= T_2(\theta, \phi) \boldsymbol{x},
    \end{aligned}
\end{equation}

With the above errors, the transmission matrix $T_2$ from Eq. \ref{eqn:mzi} becomes:
\begin{equation} \label{eqn:mzierror}
    \begin{aligned}
    \widehat{T}_2(\theta, \phi) &=  B_{\epsilon_r}(\lambda) \begin{bmatrix}
          e^{i (\theta + \delta\theta(\lambda))} & 0 \\
          0 & 1
    \end{bmatrix} B_{\epsilon_\ell} \begin{bmatrix}
          e^{i (\phi + \delta\phi(\lambda))} & 0 \\
          0 & 1
    \end{bmatrix} \\
    B_\epsilon &= \begin{bmatrix}
          C_\epsilon & iS_\epsilon \\
          iS_\epsilon & C_\epsilon
    \end{bmatrix}
    \end{aligned}
\end{equation}
where we model the beamsplitter error using $C_j = \cos\left(\frac{\pi}{4} + \delta_j\right), S_j = \sin\left(\frac{\pi}{4} + \delta_j\right)$, where $j = \ell, r$ stands for the left and right beamsplitters of the MZI.

The errors may be statistically modelled from wavelength and fabrication variations as follows:
\begin{enumerate}
    \item The coupling errors $\delta(\lambda) \sim \mathcal{N}(\mu_{\mathrm{bs}} (\lambda - \lambda_c)^2, \sigma)$, where $\mu_{\mathrm{bs}} (\lambda - \lambda_c)^2$ is a dispersion contribution with scaling $\mu_{\mathrm{bs}}$ and $\sigma$ is fabrication or drift error which is present in an MZI.
    \item The phase errors $\delta\eta(\lambda) \sim \mathcal{N}(\mu_{\eta} (\lambda - \lambda_c), \sigma_{\mathrm{ps}})$, where $\mu_\eta (\lambda - \lambda_c)$ is a dispersion contribution with scaling $\mu_{\eta}$ and $\sigma$ is crosstalk or drift error which is present in an MZI.
    \item When incorporating loss imbalance error, we have $\delta\xi(\lambda) \sim \mathcal{N}(\mu_{\mathrm{loss}}\eta_j, \sigma_{\mathrm{loss}})$. 
\end{enumerate}

Note that loss imbalance errors $\delta\xi$ only needed in the positions of the phase shifters; all remaining losses may be ``pushed'' to the end of the mesh by algorithmically applying commutations of common mode losses in the MZIs \cite{Oszmaniec2018ClassicalParticles}. Ultimately, grating coupler efficiency variations and the algorithm to move all losses to the branches will combine to give an array of $N$ independent loss elements, but this can be effectively calibrated out by the network by scaling the power threshold $p_{\mathrm{th}}$ by these constant loss terms.

\subsection{Phase shifter calibration}
As discussed in the main text, the calibration protocol involves sweeping phase shifter voltages such that the phase is calibrated from $0$ to $2\pi$.

Each phase shifter is calibrated by optical interference measurements evaluated at each MZI output in the mesh. The split transmissivity measured from spots ca be modelled as $t = \sin^2 \theta$, where $\theta$ is twice the phase shift in the internal arm, which is used for calibration:
\begin{equation}
    t = \frac{p_t}{p} \approx \frac{p_t}{p_r + p_t}
\end{equation}
where $t$ is the transmissivity, $p$ is the total power at the input, $p_t$ is the cross state grating power and $p_r$ is the bar state grating power determined by summing up pixel values from the camera.

The model is:
\begin{equation}
    \begin{aligned}
        \theta &= p_0v^3 + p_1 v^2 + p_2 v + p_3\\
        t &= a \sin \theta + b.
    \end{aligned}
\end{equation}

Empirically, it suffices to fit $v^2 = q_0 \theta^3 + q_1 \theta^2 + q_2 \theta + q_3$ to convert voltage to phase.

The calibration proceeds by first calibrating all internal phase shifters ($\theta$ phase shifts) and then calibrating all external phase shifters ($\phi$ phase shifts) using the already calibrated $\theta$ phase shifts. In order to address each of the individual MZIs, we use the calibration approaches argued in Refs. \citenum{Mower2015High-fidelityCircuits, Miller2015PerfectComponents} which treats already-calibrated MZIs as switches to progressively calibrate MZIs from input-to-output in the network. A useful trick for calibrating the external phase shifts is to create ``meta-MZIs'' in the mesh, which is a technique also used in Refs. \cite{Shen2017DeepCircuits, Harris2017QuantumProcessor, Prabhu2020AcceleratingCircuits} where external phase shifts $\phi$ can be treated as internal phase shifters in an MZI by setting appropriate $\theta$ phase shifts to $\pi / 2$.

\appendix

\section{Mining pool implementation}

It is important to have a sense for the expected hash error rate $\langle \epsilon \rangle$ for a given LightHash-based miner upon widespread adoption of a photonic proof of work-based cryptocurrency. While an accurate hash computation isn't required unless the block is solved, intermediate mining rewards in ``mining pools'' are typically offered to miners in units of ``shares'' that have a much lower difficulty $B_{\mathrm{share}} \ll B$ than the final proof of work solution, where $B$ as defined earlier represents the number of bits for proof of work. Mining pools act as a single mining entity consisting of many individual miners that earn shares in proportion to the amount of work they do. Mining pools are chosen by miners who do not have the capability of getting significant profits by mining alone. Getting a share is generally proof that the ``correct'' proof of work is being performed. Accurate mining is important to ensure that these shares can be earned at a sufficiently high percentage to reap profits. Therefore, the error rate multiplied by the number of shares will yield the total profit for the miner. Note that there is no penalty for a failed calculation for a nonce that does not yield any share reward.
\begin{figure*}
    \centering
    \includegraphics[width=\textwidth]{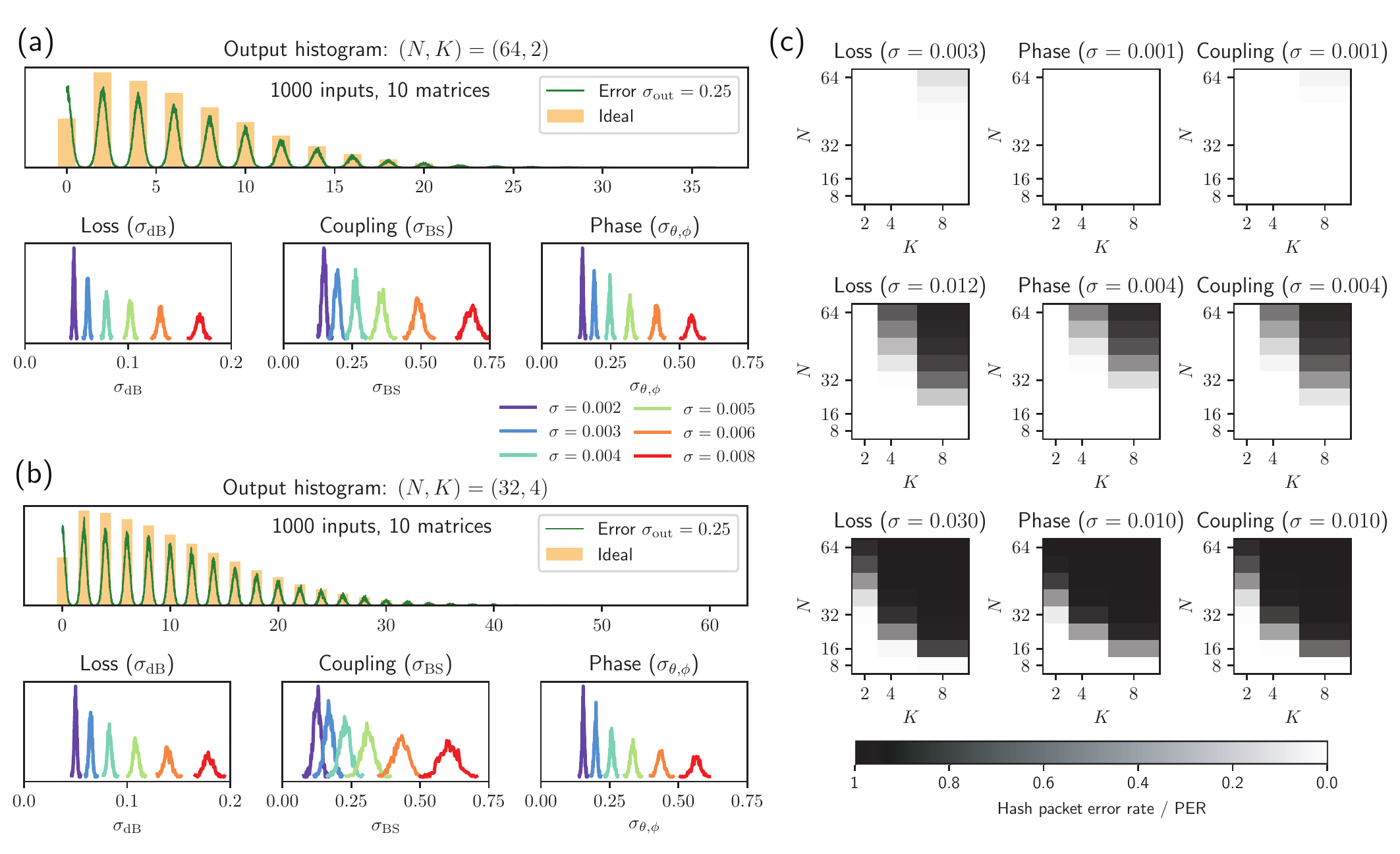}
    \caption{We extend the simulation results of Fig. \ref{fig:simulation} of the main text. Specifically, we analyze error scaling by plotting the histogram of errors and discrete outputs with $\sigma = 0.25$ for (a) $(N, K) = (32, 4)$ and (b) $(N, K) = (64, 2)$. The histograms are evaluated over $10$ matrices $Q$, $1000$ inputs $\bm{x}$ and $1280 / N$ errors given loss, coupling, and phase component errors $\sigma$. We find that the errors are similar to each other (confirming approximate $NK\sigma$ scaling) but not exactly the same (loss errors are slightly larger but coupling errors are smaller for $N = 32$). (c) We analyze the error scaling of dimension $K$ instead of component error $\sigma$, finding feasible operation of $K = 2, N = 32$ or $NK = 64$ for an error of roughly $0.01$ radians for phase and coupling and $0.03$ dB for loss. As in Fig. \ref{fig:simulation}, the transition between feasible and infeasible $(N, K)$ is very sharp.}
    \label{fig:moresimulation}
\end{figure*}

\section{General purpose usage}

The functionality of a photonic mesh as general purpose hardware is key to their fit as an optical proof of work device. For Bitcoin, the most efficient (and thus profitable) mining equipment are electronic application-specific integrated chips (ASICs) such as the Antminer S19 by Bitmain, which coerce miners to rely on the services of centralized institutions. Other coins such as Ethereum are ASIC resistant and can be profitably mined using more general purpose Graphical Processing Units (GPUs) or Field-Programmable Gate Arrays (FPGAs). This creates less hardware investment risk for miners because the GPUs and FPGAs have general use cases as opposed to mining ASICs; namely, GPUs can be used for graphics engines and machine learning applications and FPGAs are used for low-energy digital signal processing. Likewise, photonic mesh technology can be applied across several applications, with industry effort already underway in applications such as machine learning and fast signal processing. This ultimately mitigates the risk of buying new photonic hardware, if it can ultimately accrue monetary value via mining when not used for other purposes. As a simple demonstration, we show in Ref. \cite{Pai2022ExperimentallyNetworks} that our photonic mesh is capable of machine learning inference and is additionally the first to implement \textit{in situ} backpropagation training when integrated within photonic neural networks, which furthers the case for its eventual use as a cryptocurrency miner.

\section{Energy considerations}

The digital implementation of LightHash-$N, K$ consumes most of the computation if $256 \cdot  N $ floating point operations (each costing $A$ fJ of energy) exceeds the energy per SHA3-256 hash. In a fully digital implementation, if $N = 64$ and $A = 0.1$ pJ, we would have $1.6384$ nJ of energy used for matrix multiply portion. This is much more than the energy required of ASIC hash technology currently costing up to $10$ pJ per hash for Bitmain's state-of-the-art Antminer S19 device which is nearly 2 orders of magnitude lower. 

In photonic implementations, the number of electrical ops is reduced by a factor of $N / 2$ to effectively $512$ optoelectronic ops, $256$ each for the inputs and outputs. The input modulators, which may include using lithium niobate modulators \cite{Wang2018NanophotonicModulators} or silicon-organic hybrid modulators \cite{Kieninger2020Silicon-organicLoss}, each require around $1$ fJ/bit and the output comparators require at least $40$ fJ/bit \cite{Filippini2018AConversion, Miyahara2008AADCs} operating at GHz speeds. As a result, the total energy for $N = 64$ becomes roughly $10.5$ pJ of energy which is two orders of magnitude less energy than the digital and roughly the same order of magnitude as the rest of the hash. 

In summary, the guiding principle behind our energy efficient design is to perform all linear operations in the optical domain and all nonlinear or logic operations in the digital electronic domain, which leverages the core advantages of each platform. Since our protocol is dominated by linear floating point operations, most of the computation is thus handled in the optical domain, which in turn provides the strongest computational advantage possible. Of course, as explored in this work this must be evaluated in the context of systematic error, which is the principal roadblock remaining in order to realize the full potential of a photonic cryptocurrency.

\begin{figure*}
    \centering
    \includegraphics[width=\textwidth]{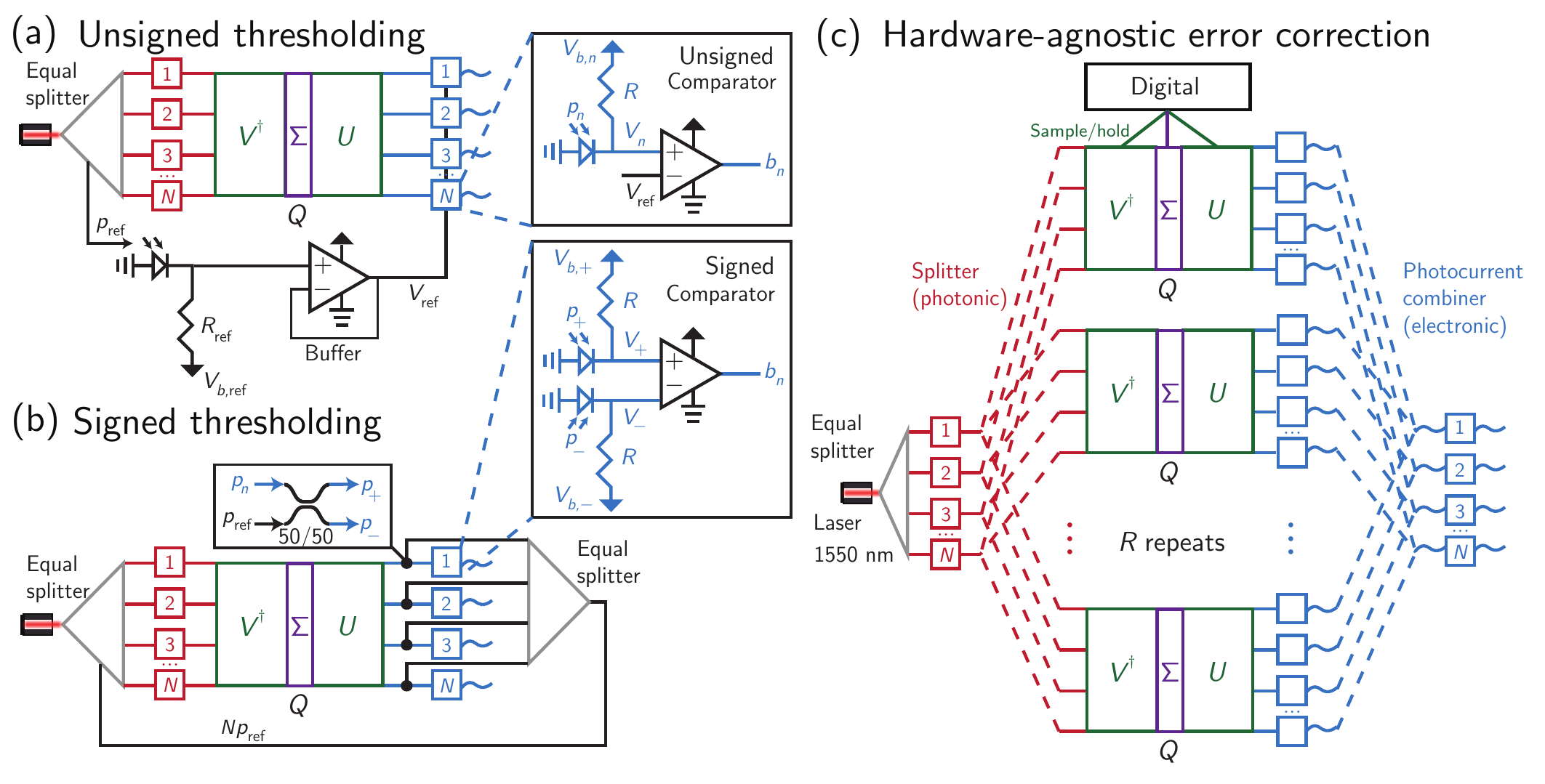}
    \caption{(a) Unsigned thresholding design can compensate for laser drift and fluctuation by using light tapped from the input. (b) Signed thresholding involves interfering a reference signal to measure the sign of the output signals. (c) A hardware-agnostic error correction SVD architecture can operate using the same energy consumption.}
    \label{fig:thresholding}
\end{figure*}

\begin{figure*}
    \centering
    \includegraphics[width=\textwidth]{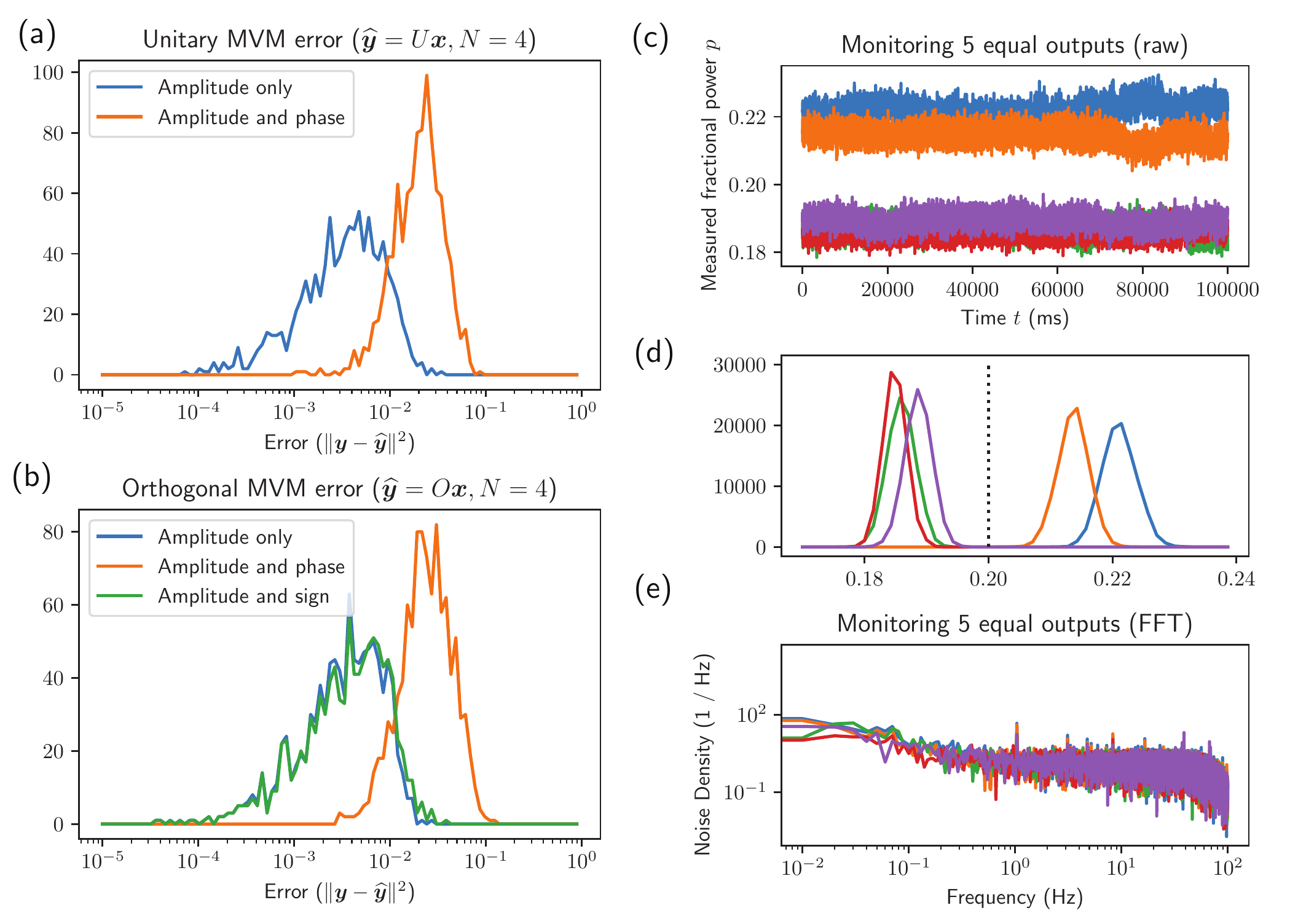}
    \caption{(a, b) We compare the unitary and orthogonal matrix multiplication errors evaluated for 1000 Haar random matrix-vector pairs. Specifically, we  compare 1. amplitude only, 2. amplitude and phase (see Ref. \cite{Pai2022ExperimentallyNetworks}), 3. amplitude and sign (for reals). We find that unitary and orthogonal matrices do not differ much import. (c, d) The expected signal is [0.2, 0.2, 0.2, 0.2, 0.2] shown using the dotted line, and biased systematic error clearly dominate compared to the standard deviation of the Gaussian distributions. (e) The FFT of the camera fractional spots (used for all optical IO, normalized over the 5 output channels). This spectrum reveals classic $1 / f$ noise (the slower variation in (c)) and white noise (broadband noise region) caused by camera photodetector noise with a noticeable dip at the end occurring at the frame rate of the camera (100 Hz).}
    \label{fig:crypto_error.pdf}
\end{figure*}

\section{Digital conversion and rescaling}

An important aspect of the LightHash algorithm is the analog-digital interface implementing the thresholding, which consumes most of the energy in our platform. Here, we explain exactly how those interfaces might be implemented in a manner that avoids any laser fluctuation or drift error contributions since the reference itself also has the same fluctuation and drift.

First, note that singular values can only be represented as lossy elements in linear optical networks. Therefore, given $Q = U \Sigma V^\dagger$ with singular values in the diagonal $\Sigma$ matrix represented as $\boldsymbol{\sigma}$, we typically divide the singular values by $\sigma_{\mathrm{max}} = \max (\boldsymbol{\sigma})$ such that no singular value exceeds 1. As a result, we need to ``remember'' this term when computing thresholds at the end of the circuit.

For the ``unsigned'' implementation of LightHash discussed in the main text and shown in Fig. \ref{fig:thresholding}(a), no additional phase reference is required. However, we still need some way to cancel out the laser noise and drift, and the implementation of the thresholding requires some more care to various scaling factors in the hardware. For laser drift compensation, one can tap out a small fraction $\xi < 1$ of the input power $P$ to establish a reference power. The corresponding input vector $\boldsymbol{x}$ is renormalized to $\widetilde{\boldsymbol{x}} = \sqrt{(1 - \xi) P} \boldsymbol{x} / \sqrt{N}$. The actual photonic network performs the operation $\widetilde{\boldsymbol{y}} = Q L \widetilde{\boldsymbol{x}} / \sigma_{\mathrm{max}}$, where $L$ is the photonic loss expressed as a fraction of the power (maximum of 1). This means that the actual output is $\widetilde{\boldsymbol{y}} = \sqrt{c_{\mathrm{out}}} \boldsymbol{y}$, where:
\begin{equation} \label{eqn:scaling}
    \begin{aligned}
    c_{\mathrm{out}} &= \frac{(1 - \xi) P L}{\sigma_{\mathrm{max}}^2 N} \\
    c_{\mathrm{ref}} &= \xi P / N
    \end{aligned}
\end{equation}
where $c_{\mathrm{ref}}$ is similar scaling factor for the reference signal split across $N$ outputs. The split reference photocurrent is related to the threshold power by $I_{\mathrm{ref}} = \eta c_{\mathrm{ref}}$ and the output photocurrents are given by $I_n = \eta \widetilde{y}_n = \eta c_{\mathrm{out}} |y_n|^2$ (Note that using the same bias voltages for all photodetectors ensures roughly same $\eta$, and additional attenuator MZIs may be used to correct for any sensitivity differences across photodetectors). Our scaling problem can therefore be reduced to finding a condition for $\xi$ such that $|y_n|^2 = p_{\mathrm{th}}$ and $I_{\mathrm{ref}} = I_n$ simultaneously, i.e. $c_{\mathrm{out}} = c_{\mathrm{ref}} / p_{\mathrm{th}}$, giving:
\begin{equation} \label{eqn:xi}
    \begin{aligned}
    \xi &= \left(1 + \frac{\sigma_{\mathrm{max}}^2}{p_{\mathrm{th}} L}\right)^{-1}
    \end{aligned}
\end{equation}
If $\xi > 1$, an alternate type of rescaling may be done by setting $\xi = 0.5$ and instead changing the bias voltages and thus the output photocurrents output by the integrated PDs to equalize the scaling:
\begin{equation} \label{eqn:eta}
    \begin{aligned}
    \frac{\eta_{\mathrm{ref}}}{\eta_{\mathrm{out}}} &= \frac{\sigma_{\mathrm{max}}^2}{p_{\mathrm{th}} L}
    \end{aligned}
\end{equation}
This rescaling calculation only needs to be performed once per block since $L, \sigma_{\mathrm{max}}^2$ only depend on the implemented $Q$ and not the inputs $\boldsymbol{x}$.

For the ``signed'' implementation of LightHash shown in Fig. \ref{fig:thresholding}(b), a circuit using a phase reference and comparator is required to measure whether the output of the mesh is a number that is positive or negative. The phase reference is generated by splitting the original input light into a reference path, which is in turn split into $N$ waveguides. The phase reference is then interfered with the overall output of the photonic network given by $\boldsymbol{y}$. In this manner we can treat the output signal $\boldsymbol{r} = \boldsymbol{i}$ (a vector of all $i$'s). Using a directional coupler to interfere any output $y$ with incoming field $i$ will give two outputs $p_+ = |y + 1|^2$ and $p_- = |y - 1|^2$. The assignment of a bit is thus given by the simple condition $p_+ > p_-$, which indicates whether a presumed real $y$ is positive or negative. Note that the aforementioned laser noise and drift will affect $\boldsymbol{r}$ and $\boldsymbol{y}$ equally since they are sourced from the same laser, so the comparator will not be affected by this error source. However, other errors such as various systematic errors in the photonic circuit will still contribute the dominant portion of the error. Note that with this technique, there is no need for rescaling based on the maximum singular value and loss as is required in the unsigned implementation.

Another point to address is whether to use an analog digital converter (ADC) to output the final aggregated bits to allow for more bits per output. An analog-to-digital converter operates at low power using a successive approximation (SAR) approach and is actually built using stages of $2^b$ comparators, where $b$ is the number of bits we use to represent the output. Compared to raw comparators, the additional overhead required for SAR might reduce the overall energy efficiency and increase latency in the overall computation per bit. This warrants future investigation because if the SAR overhead is designed to be negligible, an ADC could also be useful to aggregate bits and separate more peaks in Fig. \ref{fig:experiment} with minimal change in the hash error rate as we found using the scaling arguments in Fig. \ref{fig:simulation}(b).

Yet another alternative similar to an ADC would be to use ``parallel multithresholding'' where we split the output photonic signal into $M$ waveguides. At the cost of additional photonic loss by a factor of $M$, this split signal could be compared to multiple thresholds set between the peaks in Fig. \ref{fig:experiment}. Alternatively, we could split the photocurrent equally using a $1 \times 2^b$ splitter to measure among $2^b$ thresholds spaced 2 apart, implementing using the unsigned thresholding comparator of Fig. \ref{fig:thresholding}(a). Note that using multiple thresholds (effectively more than a single bit) will increase the error rate by a factor of at most $1 / \rho(N, K)$, at most an order of magnitude increase according to Fig. \ref{fig:simulation}.

\section{Experimental error analysis}

Chip errors in photonic meshes can be categorized as either random noise (polarization, photodetection, laser noise errors) or systematic error (loss, coupling, phase errors). In this paper, we analyze systematic error in our simulations since such errors shift the operation of the device and this are challenging to compensate straightforwardly. In this Appendix, we address collective random and systematic error contributions in our experimental setup.

Random noise can be dealt with by integrating or averaging long enough, which implies a tradeoff with latency and ultimately energy efficiency of operation. This is not the case for systematic error which is a major reason we emphasize this error type in the main text. Nevertheless, we aim to explore the various random errors in our experimental setup and the contribution of such errors to overall performance compared to systematic error.

As referenced in the main text, systematic error dominates random noise in our experimental system. This is because our random noise sources are generally straightforward to mitigate. For instance, as is shown in Fig. \ref{fig:thresholding}(a), unsigned thresholding uses a laser reference signals to compensate for any laser drift. Signal to noise ratio can also generally be improved by using longer photodetector integration time, addressing error sources such as shot noise and $1 / f$ noise caused by drift.

In our experimental setup, the photodetection noise is represented in terms of camera noise (effectively the photodetectors in our system). Camera noise consists of quantization noise in the camera pixels (14 bits of accuracy), camera photodetector shot noise, and noise due to vibration of the setup due to coupling to the mechanical stage. Other sources of noise include thermal fluctuations throughout the chip (which appears to dominate when phase shifts change) and polarization noise due to vibrations in the fiber. While not impossible to isolate these various sources of error, the systematic error in the photonic chip typically dominates these other error sources. As a consequence, we consolidate all of these errors into a single random error quantity to facilitate the comparison with systematic error. As an experimental demonstration, to confirm our claims we provide the evidence based on our results from Fig \ref{fig:crypto_error.pdf}(c, d).

In addition to characterizing random error sources, we can perform an analog comparison of expected and measured device operations, specifically matrix-vector multiplications in the real and complex domain. To do this, we use our photonic mesh to compute the dot product of the measured vector and the predicted vector of the matrix multiply $\bm{y} = U \bm{x}$ over random $U$ and random $\bm{x}$. This procedure for calculating the amplitude and phase of $4 \times 4$ matrix-vector products in our $6 \times 6$ triangular mesh is discussed in more detail in Ref. \cite{Pai2022ExperimentallyNetworks}. 

For our characterization of systematic error, we compare real (orthogonal) and complex (unitary) matrix-vector multiplication errors performed on our chip. To select a random complex $\bm{x}$, we sample from the complex normal distribution $\mathcal{N}(0, 0.5) + \mathcal{N}(0, 0.5)i$ and to select a random real $\bm{x}$, we sample from $\mathcal{N}(0, 1)$ where $\mathcal{N}(\mu, \sigma)$ represents a normal distribution with mean $\mu$ and standard deviation $\sigma$. For the complex vector, we multiply by random complex matrix $U$ sampled from the Haar measure of the unitary group. For the real vector, we multiply by random real matrix $O$ sampled from the Haar measure of the orthogonal group.

The results of this analysis are shown in Fig \ref{fig:crypto_error.pdf}(a, b), where we evaluate three types of errors: amplitude only ($\||\bm{y}| - |\widehat{\bm{y}}|\|^2$), amplitude and phase ($\|\bm{y} - \widehat{\bm{y}}\|^2$), and amplitude and sign ($\|\bm{y} - \widetilde{\bm{y}})\|^2$), where $\widetilde{\bm{y}} = |\widehat{\bm{y}}| \cdot \mathbf{sign}(\mathcal{R}(\widehat{\bm{y}}))$. A potential reason amplitude-and-phase measurements are so error-prone has to do with our readout method that relies on imperfect phase shifter calibration; this specifically affects the operation of the network for solving machine learning tasks \cite{Pai2022ExperimentallyNetworks}. For amplitude and sign error to achieve the amplitude-only accuracy, we use the phase measurement to measure only the sign and not the phase itself and use the direct output power measurements to measure the output power to ultimately minimize the error. Note that all SVD calculations in this paper assume that amplitude-only measurement is sufficient to represent error due to the $U, V$ orthogonal matrix operations on inputs $\bm{x}$, which is justified by Fig \ref{fig:crypto_error.pdf}(b).  More details on our exact implementation are provided in our Phox software \cite{Pai2022Phox:Devices} and accompanying Zenodo data availability upload \cite{Pai2022Solgaardlab/photoniccrypto:Cryptocurrency}.

\begin{algorithm}[H]
    \caption{LightHash}
    \label{alg:lighthash}
    \begin{algorithmic}[1]
        \Function{LightHash}{\texttt{blk}, $S$}
            \State $Q \gets \texttt{blk.matrix}$
            \State $D \gets \texttt{blk.difficulty}$
            \For{$s \in [1, 2, \ldots S]$} \Comment Prepare nonce
                \State \texttt{nonceList}[$s$] = $\texttt{blk.nonce} + s$
            \EndFor
            \State $N_{\mathrm{inputs}} \gets \textsc{Length}(U)$
            \State $N_{\mathrm{bits}} \gets 256 / N_{\mathrm{inputs}}$
            \State $X, \widetilde{X} \gets 0^{S \times N}$ \Comment{For batch matmul}
            \For{$s \in [0, 1, 2, \ldots S]$}
                \State $\widetilde{X}_s \gets \textsc{SHA3-256}(\texttt{blk}, \texttt{nonceList}[s])$ \Comment{Digital}
                \State $X_s \gets e^{i \pi \widehat{X}_s} / 16$ \Comment{On-chip}
            \EndFor
            \State $Y \gets QX$ \Comment{On-chip propagation}
            \State $P \gets |Y|^2$ \Comment{Photodetection}
            \State $B \gets P > p_{\mathrm{th}}(N, K)$ \Comment{Unsigned comparator}
            \For{$s \in [1, 2, \ldots S]$} \Comment{Access batch elements}
                \State $\boldsymbol{b} \gets \textsc{SHA3-256}(B_s \oplus \widetilde{X}_s)$ \Comment{Digital}
                \If{$\textsc{Int}(\boldsymbol{b}) < 2^{256 - D}$} \Comment{First $D$ bits are zero.}
                    \State \Return $\boldsymbol{b}$
                \EndIf
            \EndFor
            \State \Return $\O$
        \EndFunction
    \end{algorithmic}
\end{algorithm}

\begin{algorithm}[H]
    \caption{Optical Proof of Work}
    \label{alg:opow}
    \begin{algorithmic}[1]
        \Function{LHBlock}{\texttt{transactions}, $N$, $K$, \texttt{prevBlk}}
            \State $\mathbf{require}$ $N = 2^L \leq 256$ for $L \in \mathbb{Z}^+$ positive integer
            \State $\texttt{blk} \gets \textsc{EmptyBlock}()$
            \State $\texttt{blk.height} \gets \texttt{prevBlk}.\texttt{height} + 1$
            \State $\texttt{blk.difficulty} \gets \textsc{LHDifficulty}(\texttt{blk.height})$
            \State $\texttt{blk.prevBlkPtr} \gets \texttt{prevBlk}.\texttt{hash}$
            \State $\texttt{merkleTree} \gets \textsc{MerkleTree}(\texttt{transactions})$
            \State $\texttt{blk.merkleRoot} \gets \textsc{RootPointer}(\texttt{merkleTree})$
            \State $N_{\mathrm{blocks}} \gets 256 / N$
            \State $Q \gets O^{256 \times 256}$ \Comment{Initialize $Q$ to zeros}
            \State $n \gets 0$
            \State $\textsc{Seed}(\texttt{prevBlk}.\texttt{hash})$ \Comment{For deterministic behavior.}
            \For{$m \in [1, 2, \ldots N_{\mathrm{blocks}}]$}. \Comment{Block diagonal $Q$}
                \For{$i \in [1, 2, \ldots N]$}
                    \For{$j \in [1, 2, \ldots N]$}
                        \State $n \gets n + 1$
                        \State $\texttt{seed} \gets \texttt{merkleRoot} + n$
                        \State $q \gets \textsc{Pseudorand}([1, \ldots K], \texttt{seed})$
                        \State $Q_{m, ij} \gets 2q - K - 1$ \Comment{Block $m$, elem $i, j$.}
                    \EndFor
                \EndFor
            \EndFor
            \State $\texttt{blk.matrix} \gets Q$ \Comment{Do all chip calibration here.}
            \State $\boldsymbol{b}_{\mathrm{sol}} \gets \O$ \Comment{Solved hash}
            \While{$\boldsymbol{b}_{\mathrm{sol}}$ is $\O$}:
                \State $\texttt{blk.nonce} \gets \textsc{Pseudorand}([0, 1, \ldots 2^{N_{\mathrm{bits}}} - 1])$.
                \State $\boldsymbol{b}_{\mathrm{sol}} \gets \textsc{LightHash}(\texttt{blk}, 1)$
            \EndWhile
            \State $\texttt{blk.hash} \gets \boldsymbol{b}_{\mathrm{sol}}$
            \State \Return $\texttt{blk}$
        \EndFunction
    \end{algorithmic}
\end{algorithm}

\begin{figure*}
    \centering
    \includegraphics[width=\textwidth]{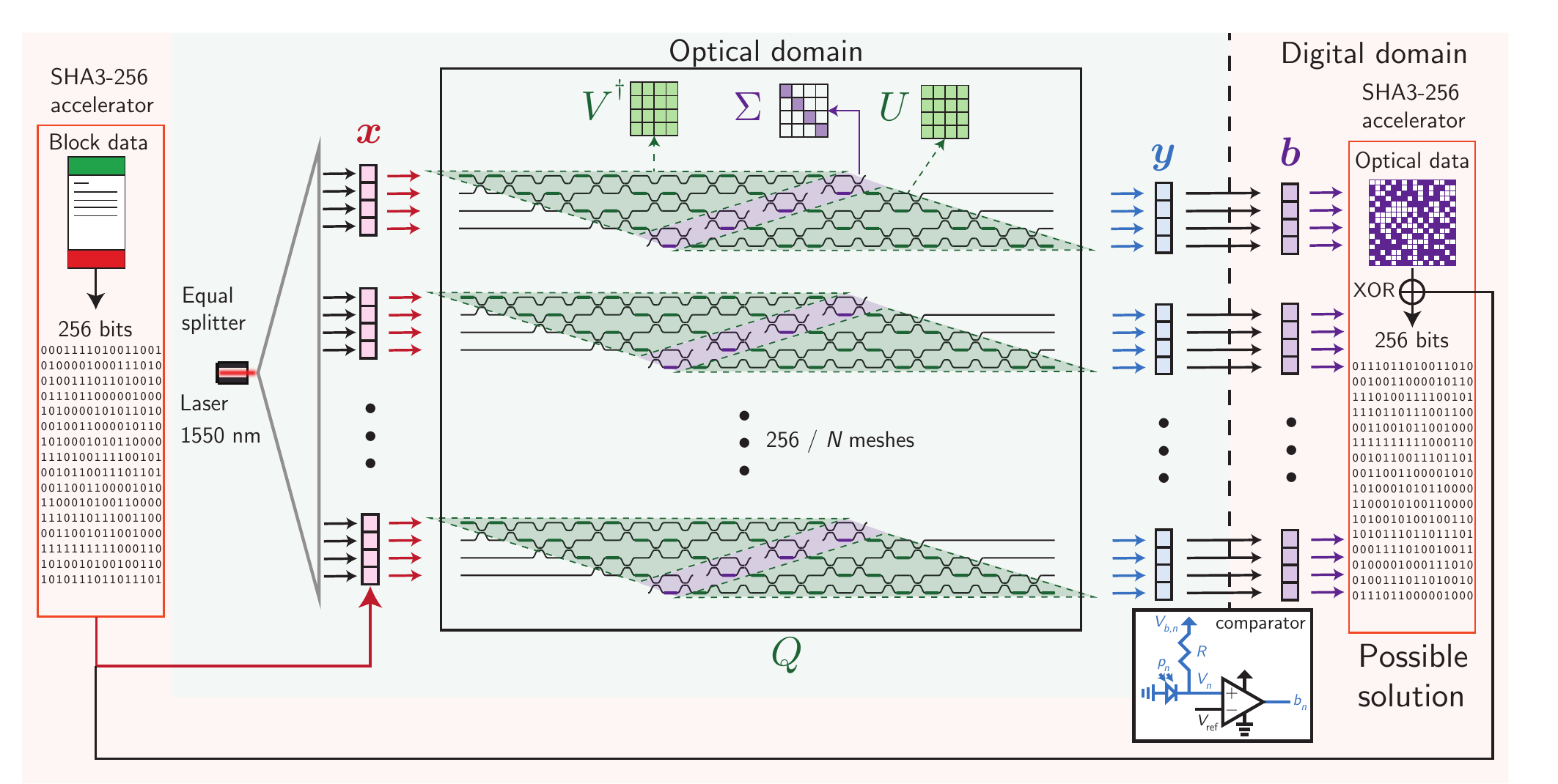}
    \caption{LightHash function flow in more detail embedded in a conceptual chip footprint with a digital processing portion (orange) and an optical processing portion (green) implementing $256 / N$ parallel SVD operations of size $N$, here shown for $N = 4$.}
    \label{fig:lighthashdetail}
\end{figure*}

\section{LightHash justification}

The LightHash function, albeit relatively simple mathematically, was chosen carefully to allow for a feasible photonic cryptographic protocol. The key insight in LightHash is that spacing possible optical output values in a discrete grid (i.e. using integer math) ultimately enables an error tolerant threshold and digital verifiability of the hash function. Additionally, LightHash has the elegant property at $K = 2$ of implementing a random walk for each vector-vector product in the overall matrix-vector product since the matrix elements and vector elements are both $-1, 1$.

The LightHash function is represented in more detail in Fig. \ref{fig:lighthashdetail}, which shows that signals are split into chunks of size $N$ and sent into $256 / N$ meshes of circuit size $N$. Unlike HeavyHash \cite{Dubrovsky2020TowardsWork}, here we provide a scheme that allows for difficulty tunability in $N$ and $K$ (numerical resolution) that allows for a feasible optical proof of work implementation.

The choice of parametrizing LightHash by $N, K$ has to do with adjusting the problem difficulty to achieve feasible bit error rate $\langle \varepsilon \rangle \leq 1\%$. We center the possible integers in the matrix to zero since LightHash is supposed to represent an optical physical random walk in discrete space. We space the integers by 2 instead of 1 to maintain integer step sizes for both odd and even $K$.

The choice of $K$ and $N$ is of course important when building any blockchain protocol around optical proof of work. Note that there exists a possible ``digital attack vector'' to consider that can ``cheat'' on LightHash based on caching results rather than performing the matrix multiply. One such issue is that we work in a finite state space of possible combinations of matrix row and vector combinations. This means that vector-vector product results can in principle be cached provided there is sufficient memory in the device. Therefore, understanding the total state space size is of critical importance in LightHash to prevent digital attacks. The system we study in this paper, $N = 4, K = 2$ to $9$, is susceptible to a digital attack vector of this kind because there are only 120 to 104976 possible vector-vector products that can feasibly be stored in RAM on a single computer. In general, there are on the order of $C := (2K)^{N}$ possible vector-vector products in the state space. Since $N, K$ affect the error equally, increasing $N$ is a more effective way to increase the overall state space to avoid this attack vector compared to increasing $K$. As an extreme example, if $K = 64, N = 2$, we have $C = 16384$ which is feasible to cache but if $N = 64, K = 2$, we have $C = 3.4 \times 10^{38}$ which is astronomically large. Despite this scaling argument, an increase in $K$ could be useful for improving the photonic advantage by requiring more bits to represent the possible outputs of the LightHash function, which would require more expensive digital hardware.

\section{LightHash pseudocode}

In this section we describe the various algorithms of required to implement LightHash and optical proof of work defined in terms of pseudocode in Algs. \ref{alg:lighthash} and \ref{alg:opow}. As mentioned in the main text, at a high level, the optical proof of work we propose is an improvement upon Bitcoin's current protocol which uses SHA256, which we have proposed as a Bitcoin Improvement Proposal \cite{Dubrovsky2021BIPPoW}. Note also our usage of the Keccak or SHA3-256 function (as opposed to other variants of SHA256) which has the following advantages: (1) it is a new-generation replacement of SHA2, developed under the NIST initiative and (2) it does not have adders and therefore results in a smaller area on chip, unlike SHA2.

There are two ways we can vary the LightHash implementation. First, we consider signed and unsigned thresholding as shown in Fig. \ref{fig:thresholding}; we choose the unsigned variety discussed in the main text. The other parameter provided in the LightHash function of Alg. \ref{alg:lighthash} is the batch size $S$ which can be used by digital implementations to parallelize the matrix multiplication across many hashes at once.

The optical proof of work protocol shown in Alg. \ref{alg:opow} considers the case $S = 1$. Using wavelength multiplexing at the expense of higher error rates, this can be extended to larger $S$ by encoding $S$ bitvectors across $S$ wavelengths propagating through the mesh simultaneously. some preliminary analysis of the bit error dependence vs wavelength is shown in the main text in Fig. \ref{fig:experiment}(f) though this may overestimate the error dispersion relation since the input vectors are also affected by the wavelength shift.

Finally, to demonstrate an implementation of LightHash, we have a bare-bones Python emulator implemented for LightHash in our Phox repository \cite{Pai2022Phox:Devices} which is also explicitly tested in our data availability repository \cite{Pai2022Solgaardlab/photoniccrypto:Cryptocurrency}.

\bibliography{crypto}

\end{document}